\documentclass[sigconf,natbib]{acmart}
\usepackage{kotex}
\usepackage{graphicx} 
\usepackage{booktabs}   
\usepackage{multirow}   
\usepackage{siunitx}    
\usepackage{amsmath}
\usepackage{enumitem}

\AtBeginDocument{%
  \providecommand\BibTeX{{%
    \normalfont B\kern-0.5em{\scshape i\kern-0.25em b}\kern-0.8em\TeX}}}

\acmConference[CIKM ’25 Paper]{34th ACM International Conference on Information and Knowledge Management}%
  {November 10--14, 2025}{COEX, Seoul, Korea}
\acmBooktitle{Proceedings of the 34th ACM International Conference on Information and Knowledge Management (CIKM ’25) -- Short Papers, November 10--14, 2025, COEX, Seoul, Korea}

\acmPrice{15.00}
\acmDOI{10.1145/3539618.3592076}
\acmISBN{978-1-4503-9408-6/23/07}

\begin{document}

\title{Improving Product Search Relevance with EAR-MP: A Solution for the CIKM 2025 AnalytiCup}

\author{JaeEun Lim}
\affiliation{
  \institution{Kangwon National University}
  \country{South Korea}}
\email{lje5370@kangwon.ac.kr}

\author{Soomin Kim}
\affiliation{
  \institution{Kangwon National University}
  \country{South Korea}}
\email{tnals5860101@kangwon.ac.kr}

\author{Jaeyong Seo}
\affiliation{
  \institution{Chonnam National University}
  \country{South Korea}}
\email{wodyd10111@jnu.ac.kr}

\author{Iori Ono}
\affiliation{
  \institution{Hiroshima University}
  \country{Japan}}
\email{m256908@hiroshima-u.ac.jp}

\author{Qimu Ran}
\affiliation{
  \institution{Hiroshima University}
  \country{Japan}}
\email{m246475@hiroshima-u.ac.jp}

\author{Jae-woong Lee}\authornote{Corresponding author}
\affiliation{
  \institution{Kangwon National University}
  \country{South Korea}}
\email{jaewoong.lee@kangwon.ac.kr}

\begin{abstract}

Multilingual e-commerce search is challenging due to linguistic diversity and the noise inherent in user-generated queries. This paper documents the solution employed by our team (EAR-MP) for the CIKM 2025 AnalytiCup, which addresses two core tasks: Query–Category (QC) relevance and Query–Item (QI) relevance. Our approach first normalizes the multilingual dataset by translating all text into English, then mitigates noise through extensive data cleaning and normalization. For model training, we build on DeBERTa-v3-large and improve performance with label smoothing, self-distillation, and dropout. In addition, we introduce task-specific upgrades—hierarchical token injection for QC and a hybrid scoring mechanism for QI. Under constrained compute, our method achieves competitive results, attaining an F1 score of 0.8796 on QC and 0.8744 on QI. These findings underscore the importance of systematic data preprocessing and tailored training strategies for building robust, resource-efficient multilingual relevance systems.


\end{abstract}





\maketitle
\section{Introduction}\label{sec:Introduction}

\subsection{Background}
E-commerce has increasingly displaced traditional shopping by offering superior convenience and has become central to daily life. The success of an e-commerce platform hinges on the quality of the user’s shopping experience, at the core of which lies product search. A search engine that retrieves the desired items quickly and accurately is pivotal for maximizing user satisfaction and, in turn, platform profitability.

Despite steady progress in improving ranking accuracy, global e-commerce introduces additional challenges. Users around the world issue queries in many languages, and the resulting linguistic barriers and cultural variation substantially complicate retrieval. Moreover, queries are typically short and noisy—often containing misspellings or non-standard expressions—which obscures user intent and makes precise matching difficult.

To address these complexities, large language models (LLMs) have emerged as a powerful alternative. With strong language understanding and broad knowledge, LLMs are reshaping core components of the search pipeline. In line with this technological trajectory, and to spur advances grounded in real industrial data, the CIKM 2025 AnalytiCup: Multilingual E-commerce Product Search Competition was organized to foster research on LLM-based multilingual product search.

\begin{figure*}[t]
\centering
\includegraphics[width=0.9\linewidth]{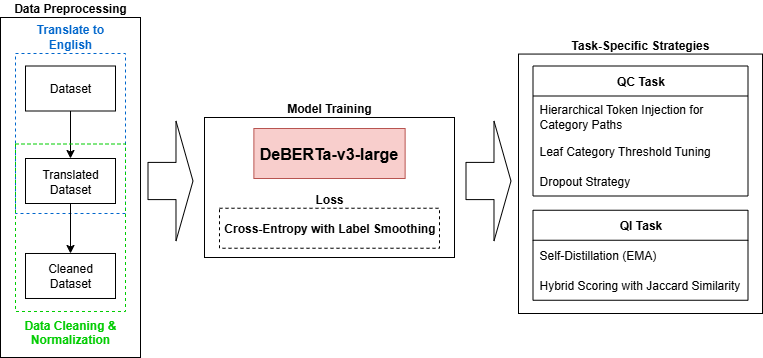}
\vspace{-3.0mm}
\caption{Overall pipeline for multilingual product search: preprocessing (translate to English, cleaning/normalization) → DeBERTa-v3-large training (label smoothing) → task-specific upgrades (QC/QI).}\label{fig:models}
\end{figure*}
\subsection{Dataset and Problem}
We study the tasks defined in the CIKM 2025 AnalytiCup using data collected from a production-scale Alibaba e-commerce platform. The competition comprises two multilingual relevance tasks.

\begin{description}[leftmargin=1.5em, labelsep=0.6em, style=nextline, font=\bfseries]
  \item[\textbullet\ Multilingual Query-Category Relevance Task (QC)]
  Product categories are organized hierarchically (e.g., Electronics $>$ Audio Devices $>$ Headphones), and each product is assigned to a leaf node. The QC task aims to infer a user’s intent from a search query (e.g., “wireless noise-canceling earbuds”) and map it to the appropriate category path. The objective is to assess whether the given category path semantically matches the query so that early-stage filtering can improve ranking precision and user experience. Field definitions and notation for queries and category paths are summarized in Table~\ref{tab:dataset_definition}.
  \item[\textbullet\ Multilingual Query-Item Relevance Task (QI)]
  Given a query–product pair, relevance is defined as whether the product satisfies the user’s expressed requirements—such as product type, brand, model, and key attributes—across all aspects. In other words, the goal is to decide whether the candidate item is exactly what the user is looking for. The meanings of the required input fields follow Table~\ref{tab:dataset_definition}.
\end{description}

\begin{table}[h]
\centering
\caption{Dataset fields and semantics used across QC and QI tasks.}\label{tab:dataset_definition}
\vspace{-3mm}
\begin{center}
{\small
\renewcommand{\arraystretch}{1} 
\begin{tabular}{c | c | c }
\toprule
Variable & Definition & Task \\
\hline
\hline
language & the language code of the query & QC, QI \\
\hline
origin\_query & the query, in the corresponding language & QC, QI \\
\hline
cate\_path & category path in English & QC \\
\hline
item\_title & the item title & QI \\
\hline
label & 1 (relevant) or 0 (irrelevant) & QC, QI \\
\bottomrule
\end{tabular}
}
\end{center}
\vspace{-4mm}
\end{table}
\begin{table}[h]
\centering
\caption{Per-language composition and data splits for the QC dataset.}\label{tab:qc_dataset}
\vspace{-3mm}
\begin{center}
{\small
\renewcommand{\arraystretch}{1} 
\begin{tabular}{c | c | c | c | c }
\toprule
Language & Code & Training Set & Dev Set & Test Set \\
\hline
\hline
English & en & 50K & 10K & 10K \\
\hline
French & fr & 50K & 10K & 10K \\
\hline
Spanish & es & 50K & 10K & 10K \\
\hline
Korean & ko & 50K & 10K & 10K \\
\hline
Portuguese & pt & 50K & 10K & 10K \\
\hline
Japanese & ja & 50K & 10K & 10K \\
\hline
Geman & de & & 10K & 10K \\
\hline
Italian & it & & 10K & 10K \\
\hline
Polish & pl & & 10K & 10K \\
\hline
Arabic & ar & & 10K & 10K \\
\bottomrule
\end{tabular}
}
\end{center}
\vspace{-4mm}
\end{table}
\begin{table}[h]
\centering
\caption{Per-language composition and data splits for the QI dataset.}\label{tab:qi_dataset}
\vspace{-3mm}
\begin{center}
{\small
\renewcommand{\arraystretch}{1} 
\begin{tabular}{c | c | c | c | c }
\toprule
Language & Code & Training Set & Dev Set & Test Set \\
\hline
\hline
English & en & 40K & 10K & 10K \\
\hline
French & fr & 40K & 5K & 10K \\
\hline
Spanish & es & 40K & 5K & 10K \\
\hline
Korean & ko & 45K & 5K & 10K \\
\hline
Portuguese & pt & 40K & 5K & 10K \\
\hline
Japanese & ja & 45K & 5K & 10K \\
\hline
Geman & de & & 5K & 10K \\
\hline
Italian & it & & 5K & 10K \\
\hline
Polish & pl & & 5K & 10K \\
\hline
Arabic & ar & & 5K & 10K \\
\hline
Thai & th & 40K & 5K & 10K \\
\hline
Vietnamese & vn & & 5K & 10K \\
\hline
Indonesian & id & & 5K & 10K \\
\bottomrule
\end{tabular}
}
\end{center}
\vspace{-5mm}
\end{table}

\begin{table*}[t!]
\centering
\caption{Summary of the Data Cleaning and Normalization Pipeline.}\label{tab:data_cleaning}
\vspace{-2mm}

\begin{center}
{\small
\renewcommand{\arraystretch}{1} 
\begin{tabular}{c | c | p{7.5cm} }
\toprule
Rule Group & Technique & Description \& Example (Before → After) \\
\hline
\hline
\multirow{3}{*}{Basic Text Normalization}
& To Lower & Converts all text to lowercase. iPhone 13 Pro → iphone 13 pro \\
& Clean Contractions & Expands English contractions to their full form. it's → it is \\
& Remove Control Char & Removes invisible characters like control characters and zero-width spaces. \\
\hline
\multirow{5}{*}{Symbol \& Character Normalization}
& Symbol Cleanup & Replaces smart quotes, dashes, etc., with standard characters. “smart quote” → "smart quote" \\
& NFKC Normalization & Normalizes Unicode compatibility/full-width characters to standard forms. ６４ＧＢ → 64GB \\
& Unicode Fractions → Decimals & Converts Unicode fraction symbols to decimals. ½ inch → 0.5 inch \\
& Normalize Range \& Mul. Symbols & Standardizes symbols for ranges and multiplication. \newline 5$\sim$10 x 5 → 5-10 * 5 \\
& Normalize Percent \& Temp & Unifies the representation of percent and temperature symbols. 23°C → 23c \\
\hline
\multirow{3}{*}{Domain-Specific Entity Normalization}
& Model Name Normalization & Unifies various expressions of major product/brand names. \newline i-phone, I phone → iphone \\
& Numbers \& Units Normalization & Removes spaces between numbers and units and converts units to lowercase. 64 GB, 5000 mAh → 64gb, 5000mah \\
& Remove Thousand Separators & Removes thousand separators from numbers. 1,024 → 1024 \\
\hline
\multirow{5}{*}{Noise Reduction}
& Remove Emoji & Removes emojis from the text. \\
& Remove Duplicate & Reduces consecutively repeated characters or spaces. \newline helllooo → hello \\
& Remove Starting/Ending Underscore & Removes underscores from the beginning/end of the text. \\
& Compress Consecutive Duplicate Tokens & Compresses consecutively duplicate tokens. new new → new \\
& Optional Removal of Promotional Phrases & Optionally removes promotional phrases like "free shipping", "best seller". \\
\bottomrule
\end{tabular}
}
\end{center}
\vspace{-4mm}
\end{table*}

Table~\ref{tab:dataset_definition} presents the core fields and their semantics used across both tasks (e.g., language code, query, category path or item title, and the binary relevance label). Building on this schema, Table~\ref{tab:qc_dataset} summarizes the language composition and split sizes for the QC task, and Table~\ref{tab:qi_dataset} does the same for the QI task.

As shown in Table~\ref{tab:qc_dataset}, the QC (Query–Category) training set comprises 300{,}000 examples—50{,}000 per language for six languages (English, French, Spanish, Korean, Portuguese, Japanese). The development and test sets add four languages unseen during training (German, Italian, Polish, Arabic), totaling 10 languages with 10{,}000 examples per language. In contrast, Table~\ref{tab:qi_dataset} shows that the QI (Query–Item) training set covers seven languages with 40{,}000–45{,}000 samples per language, while the development and test sets introduce three additional unseen languages (Thai, Vietnamese, Indonesian), yielding evaluation over all 13 languages.

This design intentionally introduces unseen languages at evaluation time, posing a core challenge that demands strong cross-lingual generalization. Both tasks are binary classification—labels are ‘1’ for relevant and ‘0’ for irrelevant—and the official metric is the average of the positive-class F1 scores across QC and QI~\cite{f1wiki}.

In this paper, we present our team EAR-MP’s approach to these tasks. We detail how systematic data cleaning and normalization, together with efficient training strategies for the \textbf{DeBERTa-v3-large} model~\cite{HeGC23}, lead to competitive performance under realistic resource constraints.
\section{METHODOLOGY}\label{sec:methodology}

\subsection{Overall Architecture}
Our methodology is designed for the characteristics of multilingual, noise-prone e-commerce search data. The end-to-end pipeline consists of three stages: (i) data preprocessing, (ii) model and training strategy, and (iii) task-specific enhancements. Figure~\ref{fig:models} presents an overview of the proposed architecture.

\subsection{Data Preprocessing}
Consistent, well-formed inputs are essential for effective model learning. To this end, we first translate all data into English to remove cross-lingual barriers, and then apply extensive cleaning and normalization to standardize formats and reduce noise.

\subsubsection{\textbf{Translate to English}}
To effectively handle a multilingual dataset with a single model, we first translate all training origin\_query and item\_title fields into English. This step normalizes semantically equivalent expressions across languages into a consistent representation, allowing the model to focus on semantic relevance rather than surface-level linguistic variation. For translation, we employ Google’s Gemini 1.5 Flash~\cite{abs-2403-05530}, which provides fast and high-quality outputs.

\subsubsection{\textbf{Data Cleaning and Normalization}}
Real-world user queries and product titles contain substantial noise (e.g., misspellings, non-standard symbols, superfluous whitespace). To systematically remove such noise and enforce consistent representations, we apply a suite of rules summarized in Table~\ref{tab:data_cleaning}. The rules are organized into four groups: (i) basic text normalization, (ii) symbol and character normalization, (iii) domain-specific entity normalization, and (iv) noise removal. These steps standardize inputs while preserving salient attributes, thereby improving downstream relevance modeling.

\subsection{Model}
We adopt DeBERTa-v3-large~\cite{HeGC23} as our base model. Its disentangled attention effectively captures subtle semantic relations between queries and product titles, while the v3 pre-training objective—Replaced Token Detection (RTD)—leverages all input tokens and yields robust representations for noisy, real-world e-commerce text. At the same time, the model retains ample capacity yet remains practical to fine-tune under our RTX 4090 ×2 setup by combining LoRA~\cite{HuSWALWWC22} with DeepSpeed~\cite{RajbhandariRRH20, RasleyRRH20}, substantially reducing memory and wall-clock costs. Overall, this choice offers a strong balance between accuracy and resource efficiency, and it is well suited to the nuanced linguistic characteristics of the e-commerce domain.

\subsection{Label Smoothing}
In standard classification training, one-hot targets (assigning probability 1 to the correct class and 0 to all others) can induce excessive confidence, which in turn encourages overfitting and degrades generalization to unseen data. Label smoothing mitigates this by slightly lowering the target probability for the correct class and redistributing the residual mass over the remaining classes as a form of regularization.

Concretely, given a smoothing parameter $\epsilon$ and the original target distribution $y_k$, we construct smoothed targets $y'_k$ as
\begin{equation}
y'_k = (1-\epsilon)\,y_k + \frac{\epsilon}{K},
\end{equation}
where $K$ is the number of classes (here, $K{=}2$). This encourages the model to retain calibrated uncertainty rather than collapsing to degenerate, overconfident predictions, thereby improving robustness and stability on out-of-distribution and previously unseen examples.

\subsection{QC Task Strategies}
Given the differing characteristics of QC and QI, we introduce task-specific enhancements to maximize performance on each task.

\subsubsection{\textbf{Hierarchical Token Injection for Category Paths}}
Transformer encoders process text as a sequence of tokens, which makes it nontrivial to capture explicit hierarchical structure (e.g., ``A $>$ B $>$ C'') present in the data. To expose this information to the model, we inject special depth markers at each level of the category path—e.g., [L1], [L2], [L3]. This hierarchical token injection enables the encoder to learn the depth of each node and its relation to higher-level categories as explicit features, encouraging structural reasoning beyond simple keyword matching.
\begin{description}[leftmargin=1.5em, font=\bfseries]
  \item[\textbullet\ {Before: Electronics,Audio,Headphones}]
  \item[\textbullet\ {After: [L1] Electronics [L2] Audio [L3] Headphones}]
\end{description}

\subsubsection{\textbf{Leaf Category Threshold Tuning}}
A single decision threshold of 0.5 in binary classification may be suboptimal when applied uniformly across heterogeneous subgroups. Motivated by the observation that each leaf category can exhibit distinct data distributions and degrees of ambiguity, we tune a category-specific threshold for QC. Using predicted probabilities on the validation set, we independently search, for each leaf category, the threshold that maximizes the F1 score over a grid from 0.30 to 0.70 (step size 0.02), and apply the selected thresholds at inference time. This per-category calibration compensates for distributional differences across leaves and yields an improvement in overall performance.

\subsubsection{\textbf{Dropout Strategy}}
To mitigate overfitting and improve generalization, we strengthen regularization by applying dropout across multiple components of the model. Dropout randomly deactivates a subset of neurons during training, reducing co-adaptation and encouraging the network to learn more robust features.
Concretely, within the Transformer encoder we set a dropout rate of 25\% on hidden layers and 10\% on attention weights. We also insert an additional 20\% dropout layer immediately before the final classifier. This configuration curbs over-specialization to the training data and yields more stable predictions at inference time.

\subsection{QI Task Strategies}

\subsubsection{\textbf{Self-Distillation (EMA)}}
To improve training stability, we adopt self-distillation, in which the current network (the “student”) learns from a more stable version of itself (the “teacher”). Classical knowledge distillation~\cite{HintonVD15} transfers knowledge from a larger, stronger teacher to a smaller, efficient student. Self-distillation adapts this idea by constructing the teacher from the model’s own past. In particular, an Exponential Moving Average (EMA) of the student parameters produces a smooth and stable teacher~\cite{TarvainenV17}. We maintain the teacher weights by updating them with a high decay rate (0.999): the teacher at step $t$ is the EMA of the student up to step $t$.
The overall training objective combines the ground-truth cross-entropy with a distillation term from the EMA teacher. The final loss $\mathcal{L}_{\text{total}}$ is
\begin{equation}
\mathcal{L}_{\text{total}} \;=\; \alpha \cdot \mathcal{L}_{\text{CE}} \;+\; (1-\alpha) \cdot \mathcal{L}_{\text{KL}},
\end{equation}
where $\mathcal{L}_{\text{CE}}$ is the standard cross-entropy against the hard labels, and $\mathcal{L}_{\text{KL}}$ encourages the student to match the teacher’s predictive distribution. We set $\alpha=0.5$ to balance the two terms, and use temperature scaling $T=2.5$ to soften both teacher and student logits prior to the KL term, which strengthens the distillation signal.

\subsubsection{\textbf{Hybrid Scoring with Jaccard Similarity}}
While language models excel at capturing contextual and semantic similarity, they can miss explicit lexical matches that are often crucial in product search. Conversely, classical text-matching methods such as the Jaccard similarity~\cite{Fletcher018} emphasize keyword overlap but do not capture semantic relatedness. To combine the strengths of both, we adopt a hybrid scoring scheme that aggregates the model’s semantic score with two lexical overlap signals. The final score is
\begin{equation}
S_{\text{final}} \;=\; w_p \cdot P_{\text{model}} \;+\; w_j \cdot \text{Jaccard}(Q, T) \;+\; w_c \cdot \text{Containment}(Q, T),
\end{equation}
where $P_{\text{model}}$ is the predicted relevance probability from the language model, $\text{Jaccard}(Q, T)$ is the ratio of the intersection to the union of query ($Q$) and title ($T$) token sets, and $\text{Containment}(Q, T)$ denotes the fraction of query tokens that appear in the title. The weights $(w_p, w_j, w_c)$ are tuned on the validation set to optimize task performance.

\begin{table}[h]
\centering
\caption{Task-Specific Training Hyperparameters for QC and QI.}\label{tab:hyperparameter}
\vspace{-2mm}
\begin{center}
{\small
\renewcommand{\arraystretch}{1} 
\begin{tabular}{c | c | c }
\toprule
Hyperparameter & Value (for QC Task) & Value (for QI Task) \\
\hline
\hline
Learning Rate & 4e-5 & 2e-5 \\
\hline
Batch Size & 32 & 16 \\
\hline
Number of Epochs & 15 & 6 \\
\hline
Warmup Ratio & 0.05 & 0.01 \\
\hline
Weight Decay & 5e-3 & 1e-4 \\
\hline
Max Sequence Length & 128 & 128 \\
\hline
LoRA Rank (r) & 8 & 8 \\
\hline
LoRA Alpha & 32 & 32 \\
\hline
LoRA Dropout & 0.1 & 0.1 \\
\hline
Label Smoothing & 0.05 & 0.05 \\
\bottomrule
\end{tabular}
}
\end{center}
\vspace{-4mm}
\end{table}
\begin{table*}[t!]
\centering
\caption{Ablation Study: Incremental Effect of Each Method on QC and QI F1}\label{tab:results}
\vspace{-2mm}
\begin{center}
{\small
\renewcommand{\arraystretch}{1} 
\begin{tabular}{c | c | c | c }
\toprule
Step & Methodology & QC F1 & QI F1 \\
\hline
\hline
1 & Baseline (DeBERTa-v3-large) & 0.8360 & 0.8204 \\
\hline
2 & + Translate to English & 0.8774 & 0.8750 \\
\hline
3 & + Data Cleaning \& Normalization & 0.8769 & 0.8743 \\
\hline
4 & + Label Smoothing & 0.8771 & 0.8744 \\
\hline
5 & + QC Task: Hierarchical Token Injection for Category Paths \& Per-Leaf Threshold Tuning \& Config-Level Dropout & 0.8796 & - \\
\hline
6 & + QI Task: Self-Distillation (EMA) \& Hybrid Scoring with Jaccard Similarity & - & 0.8744 \\
\bottomrule
\end{tabular}
}
\end{center}
\vspace{-4mm}
\end{table*}

\section{EXPERIMENT RESULTS}\label{sec:evaluation}

\subsection{Experimental Settings}

\subsubsection{Datasets}.
We conduct all experiments on the CIKM 2025 AnalytiCup dataset. For model selection, the provided training data are randomly split per language into training and validation sets with a 9:1 ratio. Performance is reported using the competition’s official metric—the F1 score~\cite{f1wiki} on the positive class (label=1)—and the final score is computed as the average of the positive-class F1 over the two tasks (QC and QI).

\subsubsection{Implementation Details}.
All experiments are implemented in PyTorch~\cite{pytorch} with Hugging Face Transformers~\cite{WolfDSCDMCRLFDS20,hf_transformers} and executed on two NVIDIA RTX~4090 GPUs. To train the large DeBERTa-v3-large model efficiently, we employ DeepSpeed with ZeRO Stage 2~\cite{RajbhandariRRH20, RasleyRRH20}, which reduces memory footprint and improves throughput.
Rather than full end-to-end fine-tuning, we adopt LoRA(Low-Rank Adaptation)~\cite{HuSWALWWC22} for parameter-efficient training: the pretrained backbone weights are kept frozen while a small number of trainable rank-decomposed adapters are inserted into the Transformer attention blocks. This approach enables faster, more stable optimization under limited compute.
Because QC and QI differ in data characteristics and objectives, we use task-specific hyperparameter configurations summarized in Table~\ref{tab:hyperparameter}.

\subsection{Experimental Results}
To quantify the contribution of each component in our data-centric pipeline, we start from a baseline model and incrementally add each method while recording the change in performance. The aggregate results are summarized in Table~\ref{tab:results}.

\begin{description}[leftmargin=1.5em, labelsep=0.6em, style=nextline, font=\bfseries]
    \item[Translate to English]    
    Translating all text into English yielded the largest performance gains, improving F1 by 4.14\% on QC and 5.46\% on QI. We attribute this to reducing multilingual variability to a single-language setting and thereby allowing the English-pretrained DeBERTa model to operate in its strongest regime.
    \item[Data Cleaning \& Normalization]
    Applying systematic data cleaning alone led to a slight drop in performance at that stage. However, when combined with label smoothing in later steps, it contributed positively by suppressing noise and stabilizing learning. This suggests that aggressive cleaning may remove some useful signal in isolation, but in conjunction with regularization it helps the model focus more effectively on salient attributes.
    \item[Label Smoothing]
    Label smoothing produced small but consistent gains in both tasks, improving QC by 0.02\% and QI by 0.01\%. This suggests that reducing overconfidence helps the model generalize slightly better.
    \item[QC Task Strategies]
    For QC, task-specific improvements—hierarchical token injection, leaf-wise threshold tuning, and strengthened dropout—produced an additional 0.25\% gain, leading to the best overall score. This indicates that fine-grained adjustments aligned with task structure are important for squeezing out further improvements.
    \item[QI Task Strategies]
    For QI, task-specific self-distillation (EMA) and hybrid scoring did not yield further gains beyond label smoothing in this ablation. We hypothesize that this is partly because label smoothing had already stabilized training; in separate experiments, these techniques showed positive effects when applied independently.
\end{description}

In summary, our experiments demonstrate that under limited compute, maximizing LLM performance is less about modifying the model itself and more about adopting a data-centric approach: deeply understanding the data, systematically cleaning it, and tailoring it to the task. In particular, unifying multilingual inputs into a single language produced the largest gains, while data cleaning, training-stability techniques, and task-specific optimizations delivered additional incremental improvements.

\vspace{1mm}
\section{Conclusion}\label{sec:conclusion}

We presented a data-centric approach developed by team EAR-MP for the multilingual e-commerce product search tasks in the CIKM~2025 AnalytiCup. To address the challenges of multilingual and noisy real-world data, we first normalized all text into English and applied a systematic cleaning and normalization pipeline. Building on DeBERTa-v3-large, we conducted efficient training under limited compute by leveraging LoRA and DeepSpeed, and further enhanced performance with label smoothing, self-distillation, and task-specific optimizations.
Our method achieved competitive results—\textbf{0.8796} F1 on QC and \textbf{0.8744} on QI. Ablation studies indicate that substantial gains can be realized without architectural complexity: careful data preprocessing and normalization alone significantly improve LLM-based relevance modeling. These findings suggest that, even without large-scale compute, a deep understanding of the data and a disciplined processing pipeline are key to building effective search systems.
For future work, we will (i) explore leveraging multimodal signals—such as product images—together with text to improve retrieval accuracy, and (ii) investigate more sophisticated training strategies that directly exploit the cross-lingual understanding capabilities of multilingual models, rather than translating all inputs into English.

\section*{Acknowledgments} 
This work was supported by the National Research Foundation of Korea(NRF) grant funded by the Korea government(MSIT) (RS-2023-00242528).

\bibliographystyle{ACM-Reference-Format}
\bibliography{references}

\end{document}